\documentclass[%
aip,
amsmath,amssymb,
reprint,%
floatfix,%
]{revtex4-2}

\usepackage{graphicx}
\usepackage{dcolumn}
\usepackage{bm}

\usepackage[T1]{fontenc}
\usepackage{mathptmx}

\usepackage{amsmath}
\usepackage{graphicx}
\usepackage{dcolumn}
\usepackage{bm}
\usepackage{hyperref}
\usepackage[capitalize]{cleveref}
\newcommand{\rhoN}{\ensuremath{\rho_{\rm N}}}
\newcommand{\kB}{\ensuremath{k_{\rm B}}}

\newcommand{\sr}{\ensuremath{s^{\rm {r}}}}
\newcommand{\pr}{\ensuremath{p^{\rm {r}}}}

\newcommand{\cvr}{\ensuremath{c_{\rm v}^{\rm{r}}}}
\newcommand{\cv}{\ensuremath{c_{\rm v}}}

\newcommand{\neff}{\ensuremath{n_{\rm eff}}}

\newcommand{\RRos}{\ensuremath{R_{\rm Ros}}}

\newcommand{\ar}{\ensuremath{a^{\rm{r}}}}

\newcommand{\deriv}[3]{\ensuremath{  \left(\dfrac{\partial #1}{\partial #2}\right)_{#3} }}
\newcommand{\flatderiv}[3]{\ensuremath{  (\partial #1/\partial #2)_{#3} }}
\newcommand{\flatrmderiv}[2]{\ensuremath{  (\rmd #1/\rmd #2) }}
\newcommand{\rmderiv}[2]{\ensuremath{ \dfrac{{\rm d} #1}{{\rm d} #2} }}
\newcommand{\rmd}{\ensuremath{ {\rm d} }}

\usepackage{xcolor}
\usepackage{soul}

\newcommand{\papertitle}{Connecting Entropy Scaling and Density Scaling}

\newcommand{\abstracttext}{It is shown that the residual entropy (entropy minus that of the ideal gas at the same temperature and density) is mostly synonymous with the independent variable of density scaling, identifying a direct link between these two approaches. The two-body residual entropy is demonstrated to not be a suitable surrogate for the total residual entropy in the gas phase. The residual entropy and the effective hardness of interaction (itself a derivative at constant residual entropy) are studied for the Lennard-Jones monomer and dimer as well as a range of rigid molecular models for carbon dioxide. It is observed that the density scaling exponent is related to the two-body interactions.}
\usepackage{cancel}
\usepackage{xcolor}
\usepackage{float}
\usepackage{booktabs}

\usepackage{zref-xr,zref-user}
\zxrsetup{verbose}
\zexternaldocument*{SI_RealNeff}

\graphicspath{{figs/}}

\begin{document}
	

%
\title{\papertitle}
%
\author{Ian H. Bell}
\email{ian.bell@nist.gov}
 \affiliation{%
Applied Chemicals and Materials Division, National Institute of Standards and Technology, Boulder, CO 80305
 }

\author{Robin Fingerhut}
\affiliation{%
	Technische Universit\"at Berlin, Thermodynamics and Process Engineering, Ernst-Reuter-Platz 1, 10587 Berlin, Germany
}

\author{Jadran Vrabec}
\affiliation{%
	Technische Universit\"at Berlin, Thermodynamics and Process Engineering, Ernst-Reuter-Platz 1, 10587 Berlin, Germany
}

\author{Lorenzo Costigliola}%
\affiliation{%
	Department of Science and Environment, Roskilde University, Postbox 260, DK-4000 Roskilde, Denmark
}%

\date{\today}

\begin{abstract}
\abstracttext
\end{abstract}

\maketitle

\section{Introduction}

Entropy scaling has been extensively studied in recent years (refer to Ref. \citenum{Dyre-JCP-2018-Review} for a review) as a means of connecting dynamics and equilibrium thermodynamics. A requirement for applying this approach to the transport properties of real fluids is a reliable model for the residual entropy (the difference between the entropy and the entropy of an ideal gas at the same temperature and density), which is usually obtained from an empirical equation of state (EOS), or more computationally costly molecular simulation methods, limiting the usefulness of entropy scaling. Representing the residual entropy straightforwardly in terms of temperature and density would therefore be appealing and broaden the range of fluids for which entropy scaling might be applied. The goal of this work is to demonstrate that this representation is less impossible than it might seem, and furthermore, it reveals a heretofore unknown link between density scaling and entropy scaling approaches in the entire phase diagram. 

\begin{figure}[H]
	\includegraphics[width=3.5in]{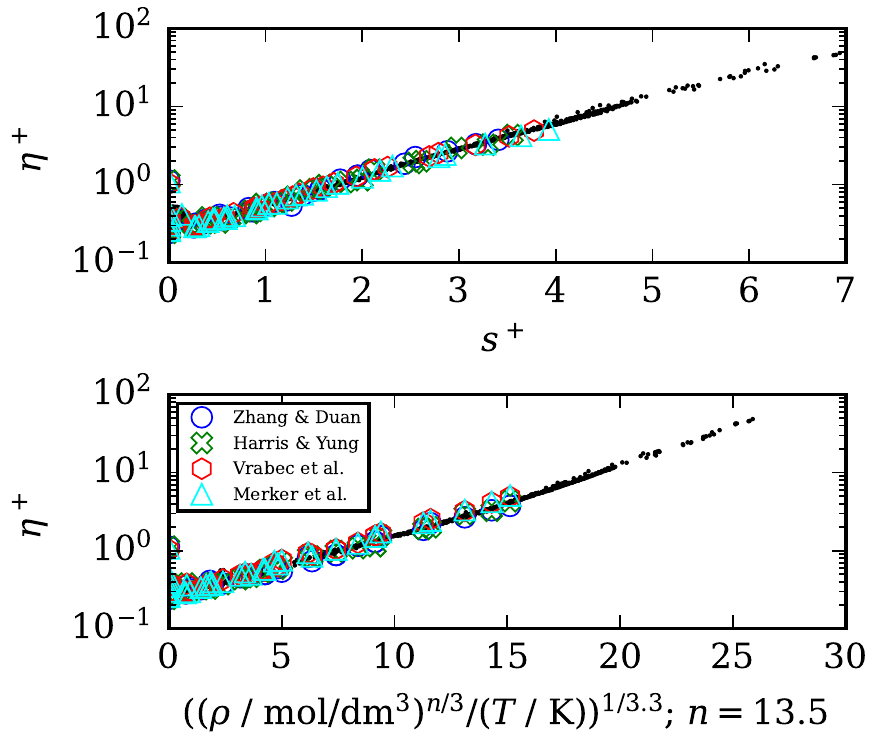}
	\caption{Modified entropy scaling applied to shear viscosity data as a function of residual entropy (top panel) and density scaling variable (bottom panel) applied to shear viscosity data for CO$_2$ (open markers: selected simulation results from this work, points: experimental data collection from Ref. \citenum{Bell-PNAS-2019}). \label{fig:CO2summary}}
\end{figure}

As a first demonstration of the motivation of this paper, we overlay shear viscosity data of CO$_2$ as a function of the independent variable of each approach in \cref{fig:CO2summary}. The dependent variable is $\eta^+=\eta/(\rhoN^{2/3}\sqrt{m\kB T})\times(s^+)^{2/3}$, where $\eta$ is the shear viscosity, $\rhoN$ is the number density, $m$ is the mass of one entity (atom or molecule), $\kB$ is Boltzmann's constant, $T$ is the temperature and $s^+$ is the reduced residual entropy defined later on. The viscosity data were scaled according to the modified entropy scaling approach introduced in Ref. \citenum{Bell-PNAS-2019}. The density scaling exponent of $13.5$ was taken from Ref. \citenum{Fragiadakis-PRE-2011}. The quantity $\eta^+$ combines macroscopic scaling (a requirement for isomorph theory), and the plus-scaling introduced in Ref. \citenum{Bell-PNAS-2019}. We will revisit each of the elements, but for now the key point is that the two approaches yield almost linear relationships in semi-log coordinates.

\subsection{Entropy Scaling}

\noindent The observation of a correlation between the variation of transport coefficients (e.g., self-diffusion coefficient or shear viscosity) of simple liquids and their residual entropy can be traced back to Rosenfeld in 1977 \cite{Rosenfeld-PRA-1977}. In Rosenfeld's work, simulation results for different systems were presented and simple exponential relations between dimensionless values of self-diffusion coefficient or shear viscosity and residual entropy were proposed. A fundamental observation was that in order to observe these correlations between transport coefficients and residual entropy, the physical quantities need to be made dimensionless by using macroscopically reduced units \cite{Rosenfeld-JPCM-1999}. In macroscopically reduced units, lengths are measured in terms of the average interparticle distance $\rhoN^{-1/3}$, with $\rhoN$ being the number density, and energies in terms of $\kB T$. A tilde above the quantity of interest will indicate in the following that the quantity is expressed in macroscopically reduced units. Similar results were later found by Dzugutov \cite{Dzugutov-NATURE-1996} which was the start of a growing interest in the entropy scaling approach. While the initial focus was on understanding the nature of the relationship between residual entropy and scaled transport properties, this approach has been shown to be applicable to a broad range of fluids (as long as they behave classically and do not form strong directional bonds). The review of Dyre \cite{Dyre-JCP-2018-Review} in 2018 summarized the progress in this field up to that date. Since then, additional studies have considered the physical basis of this approach \cite{Bell-PNAS-2019,Bell-JCED-2019-abinitio}, and applied the technique to different systems: the Lennard-Jones fluid \cite{Bell-JPCB-2019-LJ}, refrigerants \cite{Yang-IECR-2021-refrig,Yang-JCED-2021-etarefrig}, and alkanes \cite{Bell-JCED-2020-alkanes,Bell-JCED-2020-propane}.

    
\subsection{Density Scaling}
    
Density scaling is trivially valid for systems interacting via inverse-power-law (IPL) potentials of the form $V(r)\propto r^{-n}$, where $r$ is the molecular center of mass separation. For this family of systems, the dynamic and thermodynamic properties are not functions of $T$ and $\rho$ independently but depend on their combination $\Gamma=T\rho^{-n/3}$. As a consequence, the thermodynamic surface of these systems is one-dimensional. IPL fluids exhibit a one-to-one mapping between $\Gamma$ and residual entropy (see the supporting information of Ref. \citenum{Bell-PNAS-2019}) and indeed, between $\Gamma$ and all thermodynamic and transport properties with the application of an appropriate scaling. 

The density scaling approach takes a fluid of interest governed by a non-IPL potential and expresses its macroscopically scaled transport properties in the form $\widetilde{X}=f(T/\rho^{n/3})$, where $X$ is one of shear viscosity, thermal conductivity, or self-diffusion coefficient, and $n$ is in this case a fluid-specific constant for the entire phase diagram \cite{Casalini-JCP-2019}. Density scaling is thus predicated on the assumption that the effective interaction potential between molecules can be approximated by $V\propto r^{-n}$. Density scaling has been investigated for a wide range of systems, including Lennard-Jones models, modified Buckingham fluids, metals \cite{Hummel-PRB-2015}, flexible molecular analogs \cite{Galliero-JCP-2011}, and has proven to be useful also in glass-forming liquids \cite{Alba-Simionesco-JCP-2002}. A major concern for using this approach is the evidence that the density scaling coefficient $n$ has been shown to be not constant both in computer simulations \cite{Bohling-NJP-2012} and in experiments \cite{Sanz-PRL-2019, Ransom-JCP-2019}. 

    

\textit{So, how can entropy scaling and density scaling be reconciled?} This work demonstrates that the use of a constant exponent has the effect of making the unique variable of density scaling a monovariate function of residual entropy. In other words, density scaling with a constant effective hardness and entropy scaling are very close cousins.

%
%
%
%
%
%
%
%

\subsection{Isomorph Theory}

Both entropy scaling and density scaling indicate that a relation between the dynamics of fluids and their thermodynamic properties exists (see for instance \cref{fig:CO2summary}), but do not provide a satisfying explanation for \textit{why} this is the case. Rosenfeld's reference to the hard sphere system in the dense fluid phase as an explanation for the success of entropy scaling is hard to accept in gas-like phases. The assumption that repulsion, modeled as IPL interactions, yields density scaling has been challenged, both with simulations and experiments. Simulations showed that $\widetilde{X} = f(T/\rho^{n/3})$ is too simple an approximation \cite{Bohling-NJP-2012}, and experiments \cite{Sanz-PRL-2019, Ransom-JCP-2019} indicate that, in general, the density scaling exponent depends on the thermodynamic state point. A way to consistently link density scaling and entropy scaling is provided by isomorph theory, briefly introduced below. For a more complete overview of the theory we refer to Refs. \citenum{Dyre-JCP-2018-Review,Dyre-JPCM-2016,Gnan-JCP-2009-PartIV}. 

According to isomorph theory, it is possible to identify regions in the phase diagram of a given liquid in which its behavior is simpler. This region can be identified with simulations by evaluating where the correlation coefficient $R_{\rm Ros}$ \cite{Bailey-JCP-2008-PartI} is greater than $0.9$:
\begin{equation}
	\label{eq:RRos}
	R_{\rm Ros} = \frac{\left\langle\Delta U\Delta W\right\rangle}{\sqrt{ \left\langle(\Delta U)^2\right\rangle \left\langle(\Delta W)^2\right\rangle}}
\end{equation}
In the definition of $R_{\rm Ros}$, $\Delta U$ and $\Delta W$ are the deviations of the instantaneous values of potential energy and virial from its average value, respectively. The $\langle\dots\rangle$ syntax indicates the average of the argument over a canonical ensemble. The quantity $R_{\rm Ros}$ can also be evaluated with experiments under some approximations, as in Fig. 3 of Ref. \citenum{Pedersen-PRL-2008}. In the $R_{\rm Ros}>0.9$ regions, the phase diagram of the system is effectively one dimensional and the structure and dynamics of the system are invariant, when expressed in the macroscopically reduced units introduced earlier, i.e., along curves of constant residual entropy, which are called isomorphs. These invariances have been verified in several works, both with computer simulations \cite{Ingebrigtsen-PRX-2012} and experiments \cite{Hansen-NATCOMM-2018}. 

In this way, isomorph theory provides a clear link between density scaling and entropy scaling, additionally predicting the invariance of reduced structure. The weak point of this approach is that its validity is limited to some regions of the phase diagram and cannot explain, for example, the validity of entropy scaling at low densities (i.e. below the critical density).

Isomorph theory also predicts that the density scaling exponent $n/3$ depends on the thermodynamic state, as confirmed by computer simulations and experiments. The density scaling exponent is the slope of the constant residual entropy curves and can be evaluated from simulations in the canonical ensemble using the fluctuation formula \cite{Gnan-JCP-2009-PartIV}:
\begin{equation}
	\label{eq:nefffluct}
	\neff = 3\frac{\left\langle\Delta U\Delta W\right\rangle}{\left\langle(\Delta U)^2\right\rangle} = 3\deriv{\ln(T)}{\ln(\rho)}{\sr}
\end{equation}
where $\Delta U$ and $\Delta W$ have the same meanings as in \cref{eq:RRos}. This quantity can also be evaluated at any state point in experiments as shown in Ref. \citenum{Sanz-PRL-2019}.

This work will explore the link between entropy scaling and density scaling in the entire phase diagram of several fluids, i.e. both in the region of the phase diagram where isomorph theory can explain this link and close to the gas-liquid coexistence where this link is not clear. In order to clarify this issue, we consider three families of ``simple'' systems: the Lennard-Jones monomer, the Lennard-Jones dimer, and a range of molecular models for carbon dioxide. First we consider the residual entropy calculated for each system, we calculate its effective $n$, and finally, we show how the residual entropy and density scaling are connected.

%


\section{Methods}

\subsection{Thermodynamics}
In order to lay out the thermodynamics, we start with the definitions of the relevant quantities. The residual entropy $\sr$ is defined by
\begin{equation}
	\sr\equiv s(T,\rho)-s^{(\rm ig)}(T,\rho)
\end{equation}
where $s^{(\rm ig)}$ is the molar entropy of the ideal gas, and $s$ is the total molar entropy. In practice, this difference is not evaluated directly, rather the residual Helmholtz energy and its derivatives are used to obtain the residual entropy $\sr$ (e.g., see Eq. (6) of Ref. \citenum{Bell-PNAS-2019}). Furthermore, it is conceptually useful to consider rather than $\sr$ the non-dimensional term $s^+$ defined by
\begin{equation}
	s^+=-\sr/R
\end{equation}
where $R$ is the molar gas constant.  Other residual properties (residual pressure $\pr$, residual molar Helmholtz energy $\ar$, and residual isochoric molar heat capacity $\cvr$) are defined analogously:
\begin{align}
	\pr &\equiv p(T,\rho)-p^{\rm(ig)}(T,\rho) \\
	\ar &\equiv a(T,\rho)-a^{\rm(ig)}(T,\rho) \\
	\cvr &\equiv \cv(T,\rho)-\cv^{(ig)}(T)
\end{align}
The quantity $\cv^{\rm (ig)}$ has only temperature dependence, while the other ideal gas properties depend both on temperature and density.

The effective hardness $\neff$ is defined by \cite{Gnan-JCP-2009-PartIV} (identical to \cref{eq:nefffluct})
\begin{equation}
	\label{eq:neffdef}
\neff \equiv 3\deriv{\ln(T)}{\ln(\rho)}{s^{\rm r}}= 3\frac{\rho}{T}\deriv{T}{\rho}{s^{\rm r}}
\end{equation} 

After some thermodynamic manipulations \footnote{$\cv\equiv T(\partial s/\partial T)_\rho$, so we may write $\cvr\equiv T\left(\partial \sr/\partial T\right)_\rho$ or $\cvr/R\equiv -T\left(\partial s^+/\partial T\right)_\rho$. A similar starting identity of $(\partial s/\partial v)_T=(\partial p/\partial T)_v$ with $v=1/\rho$ yields the transformation of the numerator.}, the value of {\neff} from \cref{eq:neffdef} can also be written in the equivalent formulation 
\begin{equation}
	\label{eq:neffres}
	n_{\rm eff} = -3\dfrac{\rho\deriv{s^+}{\rho}{T}}{T\deriv{s^+}{T}{\rho}}
	= 3\dfrac{\dfrac{1}{\rho}\deriv{(p^{\rm r}/R)}{T}{\rho}}{c_{v}^{\rm r}/R}
\end{equation}
As will be shown later, the derivative $\flatderiv{s^+}{\rho}{T}$ is in general positive and $\flatderiv{s^+}{T}{\rho}$ is in general negative, and thus $\neff$ should be positive for the molecular systems studied here. Other systems can yield negative values of $\neff$ \cite{Costigliola-THESIS-2016}.

With the formalism of Lustig\cite{Lustig-MP-2012}, the residual Helmholtz energy derivatives can be obtained simultaneously in one molecular simulation run. In that framework, the density scaling exponent is defined by
\begin{equation}
	\label{eq:neffLustig}	
	\neff = -3\frac{\Lambda_{01}-\Lambda_{11}}{\Lambda_{20}}
\end{equation}
in which
\begin{equation}
	\Lambda_{ij} = (1/T)^i(\rho)^j\deriv{^i\partial^j(a^{\rm r}/RT)}{(1/T)^i\partial\rho^j}{}
\end{equation}

In the dilute-gas limit, where two-body interactions are fully captured by the second virial coefficient $B_2$, {\neff} is given by \cite{Bell-JCP-2020-neff}
\begin{equation}
	\label{eq:neffB2}
\lim_{\rho\to 0} \neff = -3\dfrac{T\rmderiv{B_2}{T}+B_2}{T^2\rmderiv{^2B_2}{T^2}+2T\rmderiv{B_2}{T}}
\end{equation}
which has recently been derived in terms of the pair potential for an infinite number of spatial dimensions\cite{Maimbourg-SPP-2020}, where the infinite spatial dimension limit is equivalent to the two-body limit in \cref{eq:neffB2}.

\subsection{Simulation Details}

The Lennard-Jones monomer was simulated using the RUMD software package\cite{Bailey-SP-2017-RUMD}. The potential was cut and shifted at the distance of $2.5\sigma$ and the potential parameters of $\sigma$ and $\epsilon/\kB$ were set to unity. The temperature was controlled with a Nos\'e-Hoover thermostat using $\tau=0.2$ as relaxation time. The timestep for the simulation was kept constant in macroscopic reduced units $d\tilde{t}=0.001$ and the system size was $N=1000$. The values of $\neff$ were obtained from the fluctuation formula in \cref{eq:nefffluct}.  The dependence of $\neff$ on the system size has been studied in appendix B of Ref. \citenum{Maimbourg-SPP-2020}.

For the other fluids, molecular dynamics (MD) simulations were performed solving numerically Newton's equations of motion with a fifth-order Gear predictor-corrector scheme by using the molecular simulation tool $ms$2~\cite{ms2_1,ms2_2,ms2_3,ms2_4}. All simulations were sampled in the canonical ensemble with the formalism of Lustig~\cite{Lustig-MP-2012} to calculate the Helmholtz energy derivatives with respect to density, inverse temperature as well as their combinations. Velocities were isokinetically rescaled to maintain the specified temperature. All CO$_2$ models given in Table~\ref{tab:CO2models} were simulated with $ms$2 as well as the Lennard-Jones (LJ) dimer, which was set to a fixed bond length of $\sigma$. The long-range interactions were corrected with the usual analytic mean-field equations\cite{ms2_1,ms2_2,ms2_3,ms2_4}. Radial distribution functions were sampled over the entire simulation run to calculate the two-body residual entropy, results of which are shown in the supplementary material. Chemical potential data $\mu_i$ were determined with Widom's test particle insertion method~\cite{Widom1963}. The shear viscosity was obtained by applying the Green-Kubo formalism~\cite{Green1954,Kubo1957} and the Einstein relations\cite{ms2_4} for the LJ dimer and the selected CO$_2$ models of \citeauthor{Zhang-JCP-2005}, \citeauthor{Harris-JPC-1995}, \citeauthor{Vrabec-JPCB-2001}, \citeauthor{Merker-JCP-2010}, \citeauthor{Errington-PHDTHESIS-1999} and \citeauthor{Hellmann-CPL-2014}.

The LJ dimer was studied in the temperature range $k_\mathrm{B} T/\varepsilon= 0.9$ to $100$ and density range $\rho\sigma^3=0.00017$ to $0.5$ with $N=1372$ particles, whereas for transport properties $N=4000$ were used. For that purpose, simulations were equilibrated by 100 Monte Carlo (MC) cycles and $10^5$ MD time steps. The production runs were performed for a period of $4 \cdot 10^6$ (transport: $3-5\cdot 10^7$) time steps with $\Delta t /(\sigma \sqrt{m/\varepsilon})=0.0005$ (respectively $0.001$ near the vapor-liquid equilibrium region). Intermolecular interactions were explicitly calculated up to the cutoff radius $r_{\rm c}=4\sigma$. 

Each CO$_2$ model listed in Table~\ref{tab:CO2models} was evaluated in the temperature range $T=250$ to $10,000$~K and density range $\rho=0.1$ to $25$~mol/dm$^3$ with $N=1372$ molecules and a cutoff radius of $r_c=14$~\AA ~(transport: $N=4000$, $r_c=17.5$~\AA). 200 MC cycles and $5\cdot 10^5$ MD time steps were used for equilibration and the production was performed for a period of $4 \cdot 10^6$ (for transport at least $15 \cdot 10^6$) time steps with $\Delta t=0.971$~fs (for $T=250$ to $600$~K: $\Delta t=1.942$~fs). Besides that, some phase space regions of the \citeauthor{Hellmann-CPL-2014} CO$_2$ fluid had to be simulated with different settings. At $T=10,000$~K from $\rho=9$ to $25$~mol/dm$^3$, an equilibration of 200 MC cycles and $8\cdot 10^5$ MD time steps was performed followed by a production of $8 \cdot 10^6$ time steps with $\Delta t=0.104$~fs. 

The use of the formalism of Lustig \cite{Lustig-MP-2012} to calculate all thermodynamic properties from the same simulation run yields the {\neff} values directly from its definition in \cref{eq:neffLustig} \cite{Mausbach-PRE-2018}. The molecular models considered for CO$_2$ are listed in \cref{tab:CO2models}.

\begin{table*}
	\caption{Molecular models for CO$_2$ considered in this work. \label{tab:CO2models}}
	\begin{tabular}{ccccc}
		\toprule
		Author & $N_{\rm sites}$ & site-site & quadrupole & $Q\times10^{40}$ / C~m$^2$ \\
		\midrule
	 	Murthy et al. \cite{Murthy-MP-1981} & 3 & LJ & point charges &  -12.6 \\
	    Potoff \& Siepmann \cite{Potoff-AICHEJ-2001} & 3 & LJ & point charges & -15.1  \\
		Zhang \& Duan \cite{Zhang-JCP-2005} & 3 &  LJ & point charges & -12.8 \\
		Harris \& Yung $^{\rm a}$ \cite{Harris-JPC-1995} & 3 & LJ & point charges & -13.7  \\
		M{\"o}ller \& Fischer \cite{Moeller-FPE-1994} & 2 & LJ & point quadrupole & -12.2  \\
		Vrabec et al. \cite{Vrabec-JPCB-2001} & 2 & LJ & point quadrupole & -12.7 \\
		Merker et al. \cite{Merker-JCP-2010} & 3 & LJ & point quadrupole & -13.6  \\
		Errington \cite{Errington-PHDTHESIS-1999,Potoff-MP-1999} & 3 & EXP-6 & point charges & -13.5 \\
		Hellmann \cite{Hellmann-CPL-2014} & 7 & empirical & point charges & -14.2 \\
		\bottomrule
	\end{tabular}\\
a: PM2, rigid
\end{table*}

The first four CO$_2$ models are qualitatively similar; they consist of three Lennard-Jones sites and point charges at each site. The next three models use two or three Lennard-Jones sites, along with a point quadrupole at the center of the molecule. The exception to this general approach are the models of Hellmann\cite{Hellmann-CPL-2014} and Errington \cite{Errington-PHDTHESIS-1999,Potoff-MP-1999}.  In these more advanced models, repulsion is roughly exponential in its form, and in the case of Hellmann \cite{Hellmann-CPL-2014}, empirical potentials have been fitted to each site-site interaction term, in order to match first principles calculations of the potential energy surface.

The quadrupole moment of CO$_2$ is equal to $(-14.31\pm 0.74) \times 10^{-40}$ C m$^2$, according to recent measurements of Chetty and Couling \cite{Chetty-MP-2011}, which is consistent with other recent analysis\cite{Beil-TCA-2021}. The quadrupole moment of the molecular models are given in \cref{tab:CO2models}. There is not a very strong correlation between the quadrupole moment $Q$ and the representation of the data considered in this work. The details of the evaluation of each potential are covered in the source code of \textit{potter} \cite{MIDAS-potter}. All calculations were done in SI units to ensure dimensional consistency.

%

Very precise second virial coefficients, their temperature derivatives, and values of {\neff} of these models were calculated with the approach described in Ref. \citenum{Bell-JCP-2020-neff}, with the use of the open-source \textit{potter} library and multicomplex algebra to obtain $B_2$ and its temperature derivatives simultaneously. The integrator was allowed to evaluate the integrand as many as $10^7$ times for each temperature.

\section{Residual Entropy}

The residual entropy is the independent variable of the macroscopically scaled transport properties in the entropy scaling framework and quantifies the loss of microstates of the system from intermolecular interactions. Residual entropy is a property that is not accessible experimentally, so it remains somewhat clouded in mystery. A comprehensive study of the residual entropy obtained from empirical thermodynamic models is called for.

\subsection{All-Body $s^+$}

\begin{figure}[H]
	\includegraphics[width=3.3in,page=1]{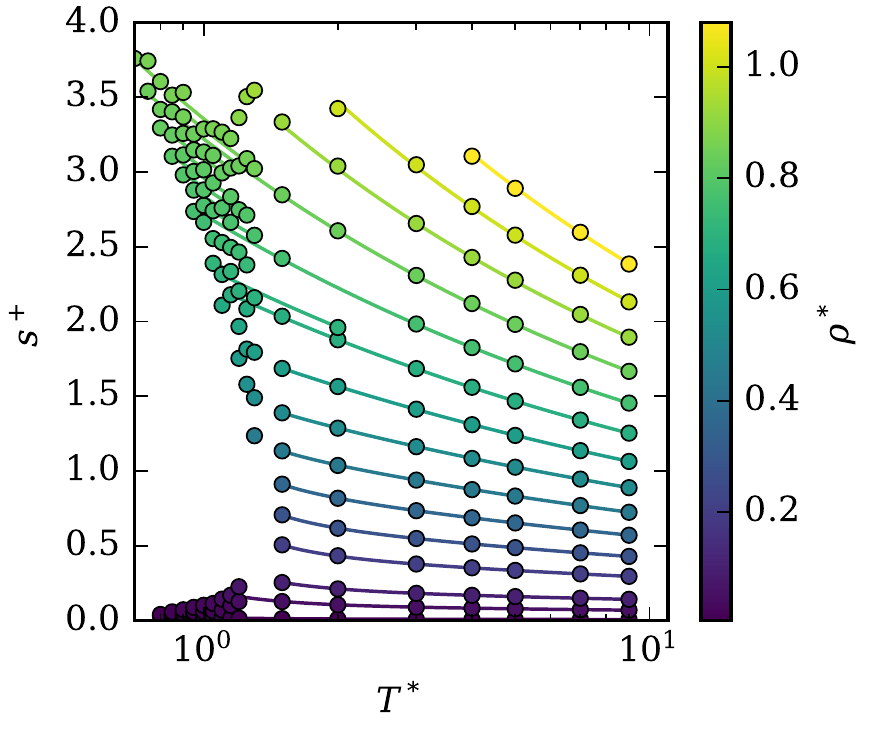}
	\caption{Values of $s^+$ for the Lennard-Jones fluid.  Markers are corrected simulation results from Ref. \citenum{Thol-JPCRD-2016} and curves are from the EOS of \citeauthor{Thol-JPCRD-2016}\cite{Thol-JPCRD-2016} \label{fig:srLJ}}
\end{figure}

For the Lennard-Jones fluid, values of $s^+$ can be obtained by thermodynamic integration \cite{Bell-NATCOMM-2020} or other sampling-based methods, and the EOS of \citeauthor{Thol-JPCRD-2016}\cite{Thol-JPCRD-2016} gives a faithful representation of this quantity. \Cref{fig:srLJ} presents the values of $s^+$ as a function of temperature and density for the Lennard-Jones fluid. 

\begin{figure}
	\includegraphics[width=3.3in,page=1]{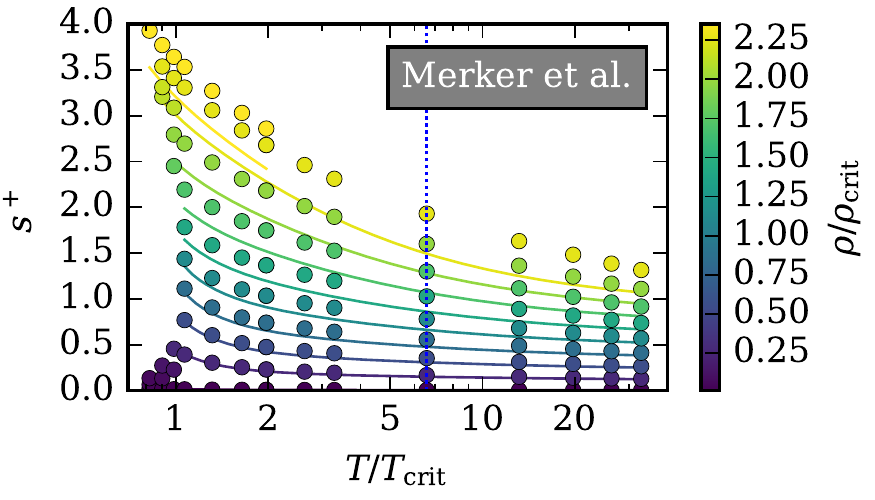}
	\includegraphics[width=3.3in,page=2]{sr_allmodels}
	\caption{Results for $s^+$ from the CO$_2$ molecular models of \citeauthor{Merker-JCP-2010}\cite{Merker-JCP-2010} and Hellmann \cite{Hellmann-CPL-2014}. The curve for each isochore is given by the Span and Wagner EOS \cite{Span-JPCRD-1996}.  The vertical dashed line indicates the temperature limit of the EOS at 2000 K. \label{fig:srMerkerHellmann}}
\end{figure}

\begin{figure}
	\includegraphics[width=3.3in,page=1]{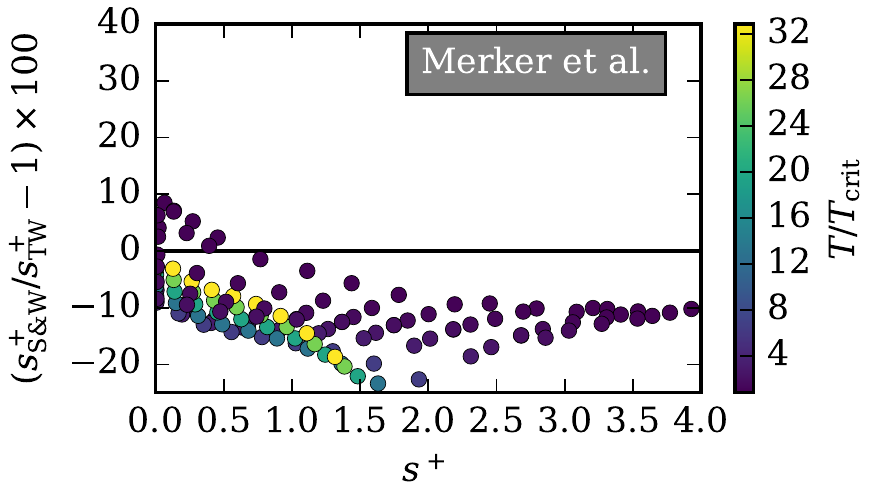}
	\includegraphics[width=3.3in,page=2]{sr_deviations_allmodels}
	\caption{Deviations of $s^+$ calculated from the CO$_2$ molecular models of \citeauthor{Merker-JCP-2010}\cite{Merker-JCP-2010} and Hellmann \cite{Hellmann-CPL-2014} (subscript TW) from the Span and Wagner EOS \cite{Span-JPCRD-1996} (subscript S\&W). \label{fig:srdevsMerkerHellmann}}
\end{figure}

In this work, we compare the values of $s^+$ obtained for the thermodynamic models for CO$_2$ with each other and with the empirical EOS of Span and Wagner \cite{Span-JPCRD-1996}. \Cref{fig:srMerkerHellmann} shows the residual entropy calculated with two molecular models, those of Hellmann and Merker et al., with the results from the Span and Wagner EOS overlaid. This result shows that the Hellmann molecular model provides a much closer agreement with the values of $s^+$ obtained from the EOS of Span and Wagner than the molecular model of Merker et al. For $T/T_{\rm crit} \lesssim 20$, it is difficult to distinguish the markers (from the molecular model of Hellmann) and the curve (from Span and Wagner). To make the comparison more quantitative, \cref{fig:srdevsMerkerHellmann} shows the deviations between the simulation data and the EOS. The other molecular models generally yield similar results to that of Merker et al., showing large deviations in residual entropy relative to the EOS. For the Hellmann model, for temperatures below the limit of the EOS at 2000 K \cite{LEMMON-RP10}, the mean absolute relative percentage error (MAPE) in $s^+$ is 2.1\%. One distinguishing feature of the Hellmann model is its representation of the effective hardness $\neff$, as shown in \cref{sec:neff}. The differences are already evident at the level of classical calculations based upon the second virial coefficient. 

For a state point either above the critical temperature, or in the gaseous phase for subcritical temperatures, scaled residual entropy at a given state point can be obtained by an integral taken at constant temperature
\begin{equation}
	\label{eq:splusint}
	s^+ = \int_0^{\rho} \deriv{s^+}{\rho}{T}{\rmd }\rho
\end{equation}
where $s^+$ in the zero density limit (that of the ideal gas) is zero. This is the typical ``thermodynamic integration'' approach familiar to molecular simulation practitioners, formulated in a different fashion. An alternative (and thermodynamically identical) representation of \cref{eq:splusint} is
\begin{equation}
	\label{eq:splusintdprdT}
	s^+ = \int_0^{\rho} \frac{1}{\rho^2}\deriv{(\pr/R)}{T}{\rho}{\rmd }\rho
\end{equation}
The formulation in \cref{eq:splusintdprdT} highlights the importance of high quality densimetry data (measurements of density $\rho$ as a function of temperature and pressure) for the representation of residual entropy. If the temperature and density dependence of pressure is well captured by laboratory measurements, the derivative \flatderiv{\pr}{T}{\rho} will also be, and the residual entropy obtained from a highly accurate empirical model fitted to these data will also be accurate.  In the case of CO$_2$ therefore, we may reasonably assume that the residual entropy obtained from the Span and Wagner EOS is probably correct within its range of validity given the large quantity of high quality densimetry data, and the excellent agreement of this EOS with these data \cite{Span-JPCRD-1996}.

\begin{figure}[H]
	\includegraphics[width=3.5in]{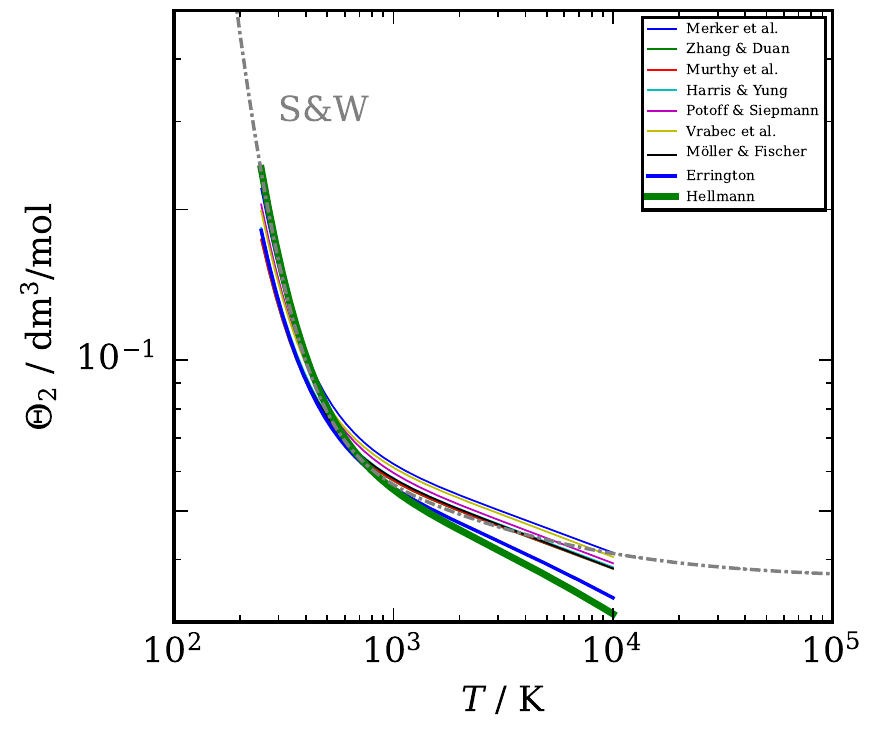}
	\caption{Values of $\Theta_2=B_2 + T\flatrmderiv{B_2}{T}$ from molecular models and from the Span and Wagner \cite{Span-JPCRD-1996} (S\&W) EOS for CO$_2$. \label{fig:Theta2_CO2}  }
\end{figure}

At low density, $s^+$ is governed by the leading term of the virial expansion as explained in Section \zref{sec:appvirialsplus} in the supporting information:
\begin{equation}
	\label{eq:splusvirial}
	\lim_{\rho\to 0}s^+ = \left(B_2 + T\rmderiv{B_2}{T}\right)\rhoN
\end{equation}
so the behavior of $\Theta_2=B_2 + T\flatrmderiv{B_2}{T}$ can provide information on the quality of the molecular model and that of the EOS. \Cref{fig:Theta2_CO2} shows the obtained values of $B_2+T\flatrmderiv{B_2}{T}$ for each model and the EOS. The model values were obtained classically with \textit{potter}. The quantity $\Theta_2$ must be positive for all temperatures because the entropy must be less than that of an ideal gas at the same temperature and density, a constraint fulfilled by all molecular models and the EOS, but the qualitative behavior of the EOS is incorrect (compared with the Hellmann model) above approximately 1000 K. The reproduction of dilute gas residual entropy values in the low-density gas (considering values of $\Theta_2$) is thus shown to be a sensitive test for the residual entropy.

%

\subsection{Two-Body $s^+_2$}
The all-body residual entropy is an inconvenient quantity to access when desiring to apply entropy scaling. In simulations, insertion methods \cite{Nezhad-MP-2016} or thermodynamic integration \cite{Bell-NATCOMM-2020} are required to obtain the residual entropy. In the real world, an equation of state is required, which is not always available for many classes of fluids. A convenient estimation of the total $s^+$ is to consider just the two-body part, a quantity that is accessible from the radial distribution function and can be straightforwardly obtained in simulation.

The total excess entropy can be given as the summation \cite{Scheiner-JCP-2021}
\begin{equation}
	s^+ = \sum_{n=2}^\infty s^+_n.
\end{equation}
It is commonly assumed in the entropy scaling literature that the two-body residual entropy $s_2^+$ is a sufficiently good representation of the total $s^+$ (Ref. \citenum{Scheiner-JCP-2021}). Where isomorph theory is valid (especially in the liquid phase), the same scaling that applies for $s^+$ should necessarily apply for $s_2^+$ (Refs. \citenum{Costigliola-THESIS-2016,Gnan-JCP-2009-PartIV}). In the gas phase, entropy scaling still works well, although by rights it should not. \textit{So, it is sufficient to use the $s^+_2$ in place of $s^+$ in general? }
	
As a test we have calculated the two-body residual entropy for each of the models for CO$_2$. To do so, the radial distribution function (RDF) $g_{ij}$ data for each atom pair was obtained for all CO$_2$ molecular models, and then integrated to obtain the $s_2^+\equiv-s_2/\kB$ via
\begin{equation}
s^+_2 = 2\pi\rho_{\rm atom}\sum_{ij}x_ix_j\int_0^\infty (g_{ij}(r)\ln(g_{ij}(r))-g_{ij}(r)+1)r^2\rmd r
\end{equation}
where $x_i$ is the mole fraction of the atom, with $i$ or $j$ being either C or O. A subtle but important point about the two-body residual entropy of molecular systems is that $\rho_{\rm atom}$ is the number density of \textit{atoms} (so atoms per volume), not the number density of sites or molecules. For molecular models where the number of sites matches the number of atoms (CO$_2$ has three atoms), the distinction is moot, but some CO$_2$ models have a number of sites that does not match the number of atoms: M{\"o}ller \& Fischer \cite{Moeller-FPE-1994} (two sites), Vrabec et al. \cite{Vrabec-JPCB-2001} (two sites), Hellmann \cite{Hellmann-CPL-2014} (seven sites).  This site/atom mismatch causes the two-body residual entropy to yield significantly different values. 

\begin{figure}[H]
	\includegraphics[width=3.5in]{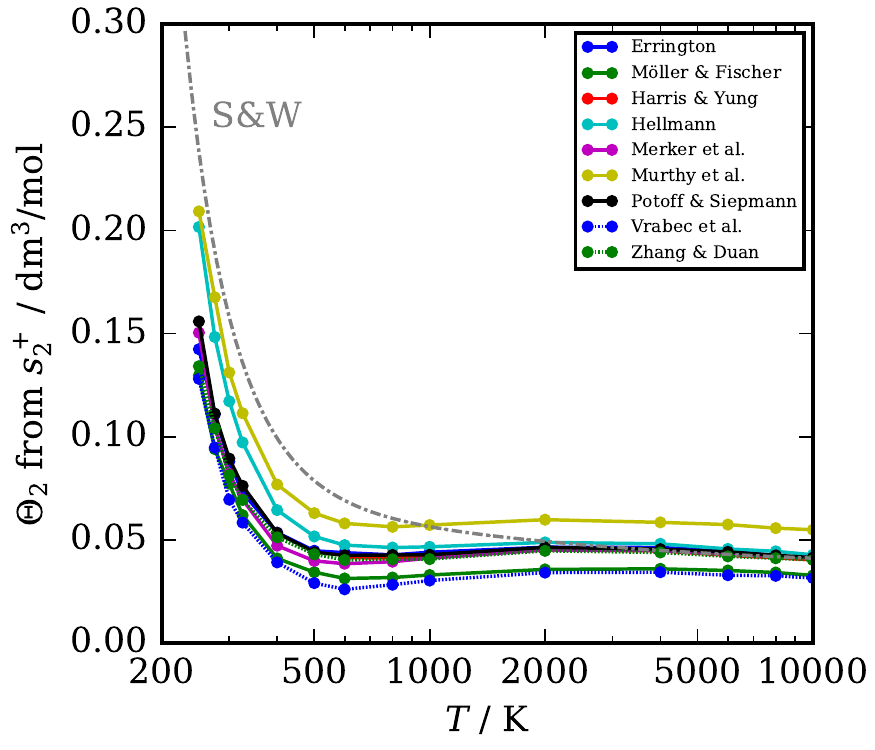}
	\caption{Values of $\Theta_2$ obtained from calculations of $s_2^+$ as a function of temperature. \label{fig:Theta2_s2vals_dilute} }
\end{figure}

Another mysterious feature of the two-body $s^+_2$ is that its dilute-gas limit does \textit{not} match that of the total $s^+$ in general. Mathematically, 
\begin{equation}
	\lim_{\rho \to 0} s^+_2 \neq \left(B_2 + T\rmderiv{B_2}{T}\right)\rhoN
\end{equation}
This result is perplexing because $s_2$ is about two-body effects, and so is the second virial coefficient.  Therefore, it stands to reason that the dilute-gas limit of $s^+_2$ should converge to that of $s^+$, but that is not the case. Values of $s^+_2$ in the gas phase were obtained in this study, and the derivative
\begin{equation}
	\Theta_{2,s^+_2} = \frac{s^+_2(\rho_{\rm min})-0}{\rho_{\rm min}-0}
\end{equation}
is the effective two-body version of $\Theta_2$ obtained by finite differentiation, where $\rho_{\rm min}$ is the lowest density at which the calculation was carried out. The values are plotted in \cref{fig:Theta2_s2vals_dilute}, from which it appears that the infinite temperature limit is the same as $\Theta_2$ for the total $s^+$.

\section{Density Scaling and Entropy Scaling}

\begin{figure}[H]
	\includegraphics[width=3.5in]{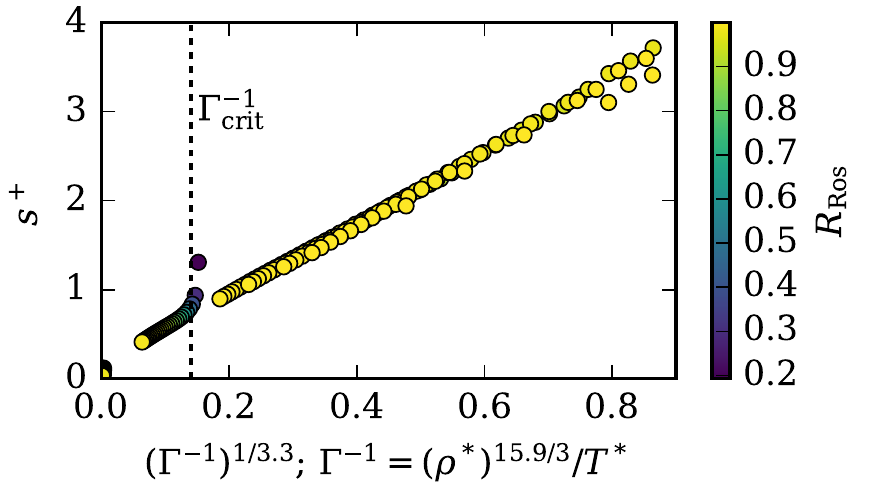}
	\caption{Scaled values of $s^+$ for the Lennard-Jones monomer compared with density scaling values for $T^* < 20$.  \label{fig:Thol_density_scaling}}
\end{figure}

To begin our comparison between density and entropy scaling, we follow the approach taken in density scaling: the use of a constant $n$ for the entire phase diagram in the definition of $\Gamma$. In the case of the Lennard-Jones monomer fluid, there is a particular value of $n$ which maximizes the correlation between $(\rho^*)^{n/3}/T^*$ and $s^+$. With this optimized value, the Spearman correlation coefficient between $(\rho^*)^{n/3}/T^*$ and $s^+$ is greater than 0.999, which represents nearly one-to-one relationship. The MD data also consider the gaseous phase and the critical region so that most of the phase diagram is covered. In order to assist with the visualization, the value of $\Gamma^{-1}$ was scaled with an exponent to linearize the relationship between $\Gamma^{-1}$ and $s^+$ in \cref{fig:Thol_density_scaling}.  This figure demonstrates that in the case of the Lennard-Jones monomer, $s^+$ and $\Gamma^{-1}$ are directly connected to each other. The particular surprise in this figure is that the relation between $\Gamma$ and $s^+$ holds even in parts of the phase diagram where $R_{\rm Ros}\ll 0.9$. In the dilute-gas limit, this scaling should break down because the leading term from the virial expansion is defined as in \cref{eq:splusvirial} which does not follow the same scaling.

\begin{figure}[H]
	\includegraphics[width=3.5in]{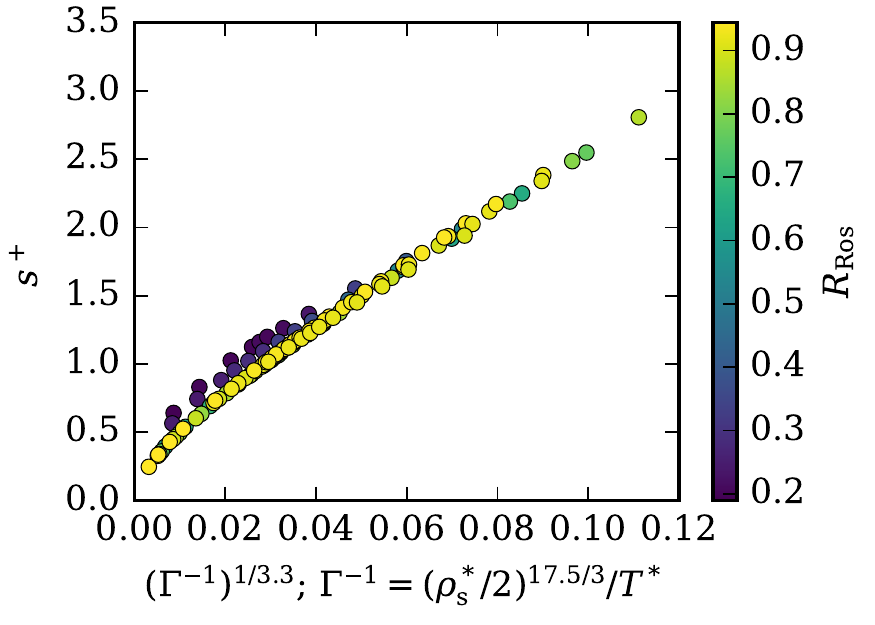}
	\caption{Values of $s^+$ for the Lennard-Jones dimer from the present simulations compared with density scaling values. Note that the density $\rho_{\rm s}^*$ is the reduced monomer number density. \label{fig:2CLJ_density_splus}}
\end{figure}

For the Lennard-Jones dimer, the qualitative picture is similar, as shown in \cref{fig:2CLJ_density_splus}. Again, a constant value of $n$ was selected that maximized the Spearman correlation between $(\rho^*_{\rm s}/2)^{n/3}/T^*$ and $s^+$. The addition of the bond to form a linear molecule does not appear to alter the core conclusion that a fixed value of $n$ is needed to form a one-to-one relationship between $(\rho^*)^{n/3}/T^*$ and $s^+$. For some of the state points with $R_{\rm Ros} < 0.3$, indicating a breakdown of isomorph theory, the mapping between the variables is slightly less strong, but aside from these deviating points, the mapping is nearly as one-to-one as for the Lennard-Jones monomer.  The values of $s^+$ for the dimer are approximately two times larger than those of the monomer at the same monomer density and temperature (microstates are removed by the fixed bond, limiting the accessible phase space). The state point dependence is shown in Fig. \zref{fig:dimerization_entropy_change} in the supporting information. 

\begin{figure}[H]
	\includegraphics[width=3.5in]{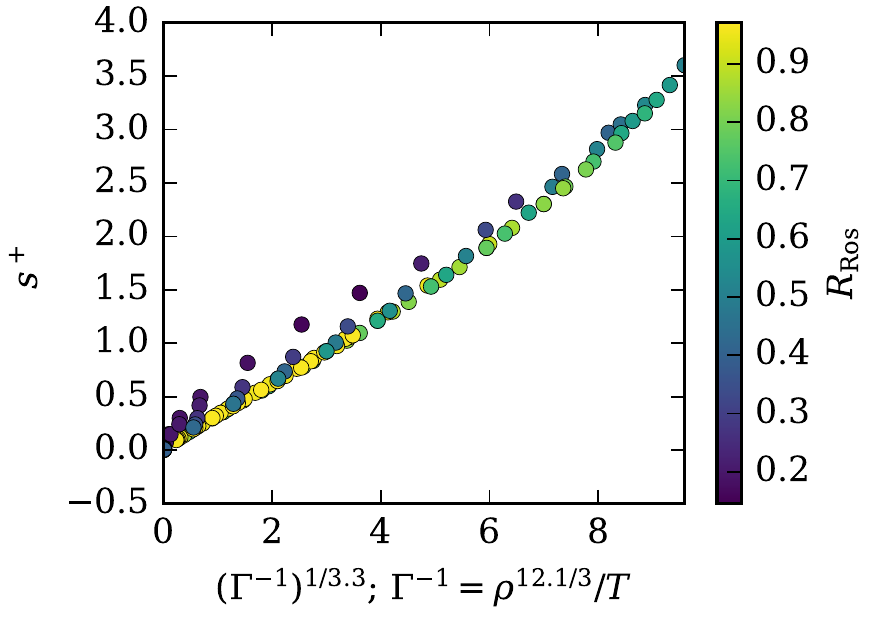}
	\caption{Values of $s^+$ from the Hellmann\cite{Hellmann-CPL-2014} model for CO$_2$ from the present simulations as a function of the density scaling values. Density $\rho$ is in units of mol/dm$^3$ and temperature $T$ is in units of K. \label{fig:Hellmann_density_splus}}
\end{figure}

For the CO$_2$ model of Hellmann\cite{Hellmann-CPL-2014}, the behavior is much the same as for the Lennard-Jones dimer. Again, $n$ was selected to maximize the Spearman correlation between $\rho^{n/3}/T$ and $s^+$ for the points with $\RRos>0.5$. \Cref{fig:Hellmann_density_splus} shows the same type of plot, but with one striking difference.  The relationship between $s^+$ and $(\Gamma^{-1})^{1/3.3}$ is qualitatively different. The curvature is convex in the case of CO$_2$ and concave in the case of the Lennard-Jones monomer and dimer.

%
%

\section{Effective hardness}
\label{sec:neff}

The analysis above primarily focused on \textit{post hoc} analysis of simulation data in order to determine the optimal value of $n$ for a particular system. \textit{What if the optimal value of $n$ is unknown?} A first glimpse of a predictive model for the optimal $n$ comes from a consideration of the effective hardness of interaction $\neff$. The quantity $\neff$ entered the vocabulary of thermodynamics with the advent of isomorph theory. The effective hardness can be conceptually thought of as the effective repulsiveness of the interactions between molecules\cite{Bailey-JCP-2008-PartII}. This intuitive understanding is somewhat incomplete, but gives a flavor.
%
%

\subsection{Lennard-Jones monomer}

\begin{figure}[H]
	\includegraphics[width=3.5in]{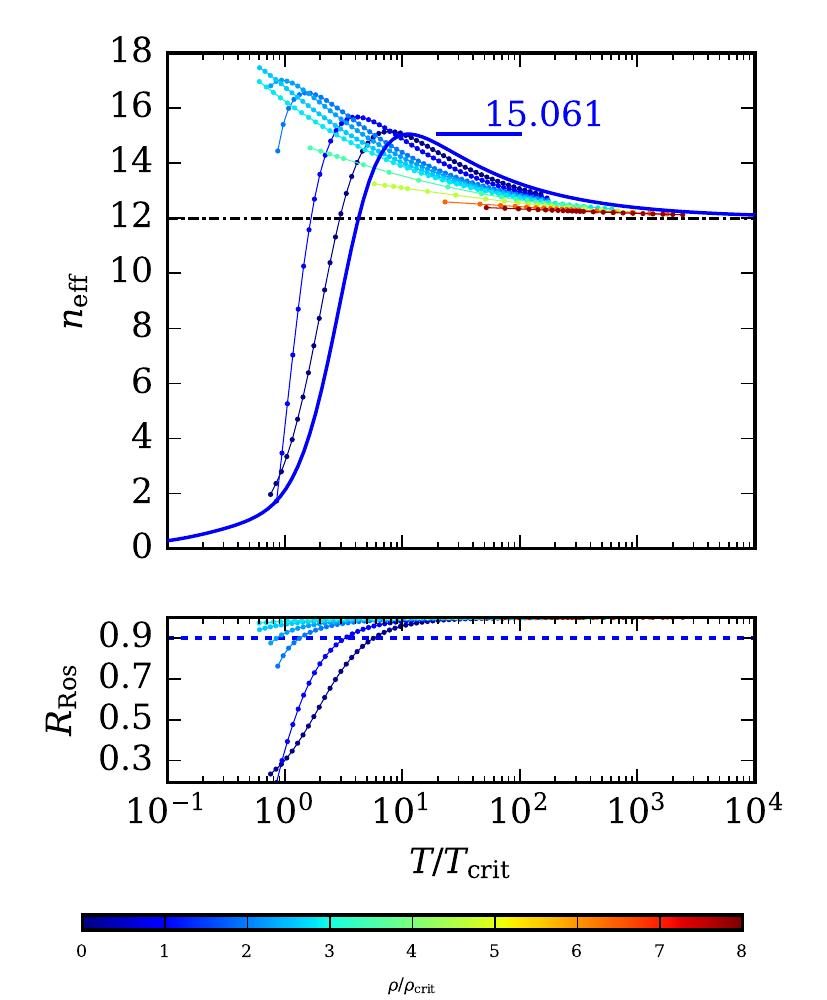}
	\caption{Values of $\neff$ and $\RRos$ for the Lennard-Jones monomer fluid. The thick curve is the value obtained from the closed form solution for the second virial coefficient published by Sadus \cite{Sadus-JCP-2018,Sadus-JCP-2019-erratumBMie}, and other values were calculated from NVT simulations performed with the RUMD software package \cite{Bailey-SP-2017-RUMD}. \label{fig:LJ_neff_Trho}}
\end{figure}

We first consider the density scaling exponent {\neff} obtained from MD simulations for the Lennard-Jones fluid. The results of these simulations are shown in \cref{fig:LJ_neff_Trho}. The calculations extend from the dilute gas up to extremely high temperatures and very dense liquid states. The dilute-gas values obtained from the second virial coefficient \cite{Bell-JCP-2020-neff} are also shown, highlighting that the values approach 12 in the infinite temperature limit. In this high-temperature limit, the interactions are entirely governed by the repulsive contribution (which is proportional to $r^{-12}$ for the Lennard-Jones fluid). For densities and temperatures more aligned with engineering applications, {\neff} has both temperature and density dependence.  Along the critical isotherm, the values of {\neff} vary from approximately 16 to zero (going towards zero at the critical point); {\neff} is decidedly not constant for even simple systems like the Lennard-Jones fluid.

The integration from \cref{eq:splusint} may equivalently be written in terms of $\neff$ as
\begin{equation}
	\label{eq:splusintneff}
	s^+ = \frac{1}{3}\int_0^{\rho} \frac{(\cvr/R)\neff}{\rho}{\rmd }\rho
\end{equation}
This expression provides a useful way of thinking about the relationship between residual entropy and $\neff$. The conceptual lesson of \cref{eq:splusintneff} is that if $\neff$ and $\cvr$ obtained from an EOS or molecular model are both correct, the residual entropy will also be. Conversely, if the values of $s^+$ are thought to be correct, and the $\neff$ is correct, the isochoric heat capacity should also be correct. However, experimental measurements of heat capacities for fluids are often characterized by relatively large experimental uncertainties, inconsistency, and in many cases by a complete lack in the open literature.

At temperatures well above the critical temperature, $\neff$ depends only relatively weakly on density \cite{Maimbourg-SPP-2020} and the representation of the residual entropy is therefore largely governed by the dilute-gas $\neff$. For instance in \cref{fig:LJ_neff_Trho}, for $T > 10T_{\rm crit}$ the variation of $\neff$ is within roughly $30\%$ of the infinite temperature limit of 12. This is why density scaling with a constant $n$ works reasonably well when studying a narrow region in the liquid region of the phase diagram, but not otherwise. Following \cref{eq:splusintneff}, constraining or obtaining the correct dilute-gas value for $\neff$ also constrains much of the liquid phase residual entropy. For liquid states, most of the variation in $\neff$ for the Lennard-Jones monomer corresponds to the region close to the critical point.


\subsection{Lennard-Jones dimer}

\begin{figure}[H]
	\includegraphics[width=3.5in]{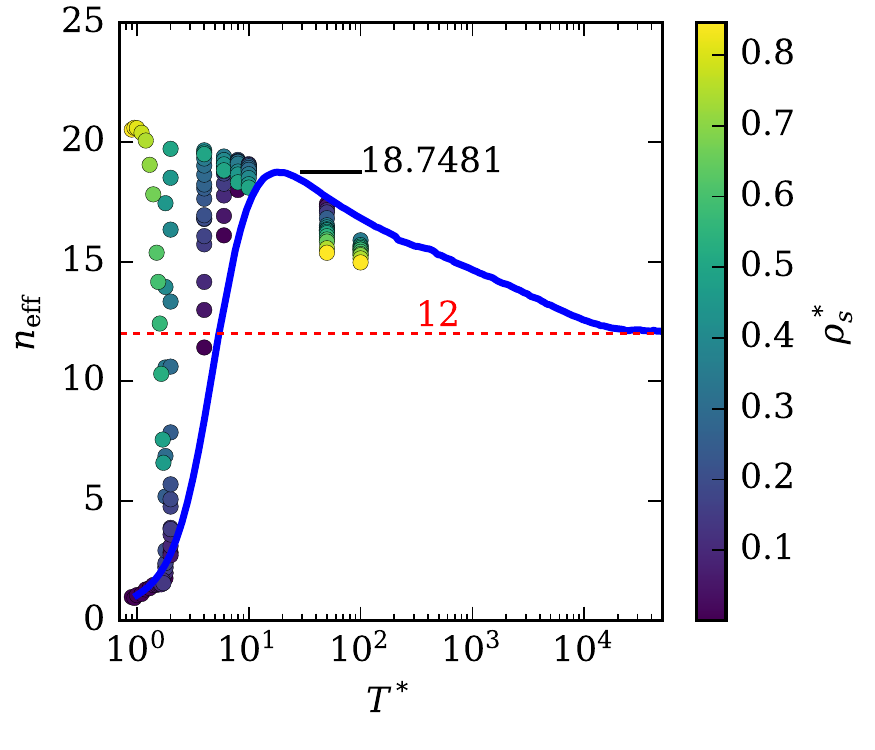}
	\caption{Values of {\neff} for the Lennard-Jones dimer fluid as a function of temperature and monomer density $\rho^*_s$ (which is twice the reduced molecular density). The dilute-gas limit (solid curve) was taken from Ref. \citenum{Bell-JCP-2020-neff} and the markers are the simulation data from this work. \label{fig:2CLJ_neff}}
	
\end{figure}

Ref. \citenum{Bell-JCP-2020-neff} considered the {\neff} in the dilute-gas limit for rigid linear chains with Lennard-Jones sites. The values of {\neff} in this work were obtained with a similar method (integration with \textit{potter} over three angles and center-of-mass separation), and are shown in \cref{fig:2CLJ_neff}. The fundamental difference between the Lennard-Jones monomer and dimer is only one of magnitude; the qualitative behavior is similar, and the vertical axis is mostly just scaled.  In the infinite-temperature limit, the value of $\neff$ also approaches 12 because at sufficiently high temperatures the dominant interaction is the pairwise repulsion of two sites governed by an $r^{-12}$ interaction \cite{Bell-JCP-2020-neff}.

\subsection{Carbon dioxide}

\begin{figure}[H]
	\includegraphics[width=3.5in]{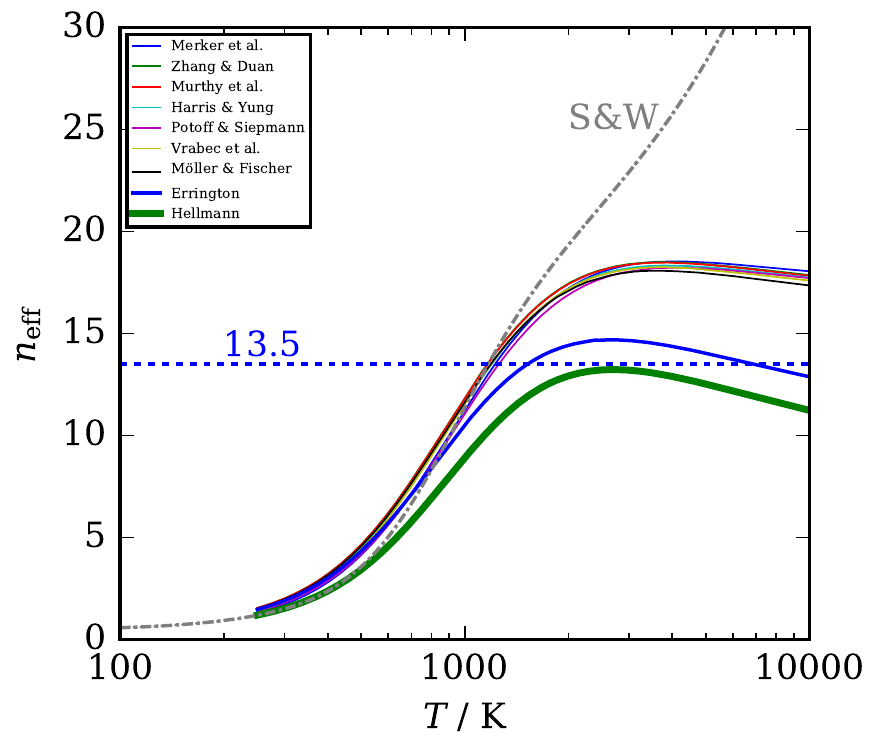}
	\caption{Values of $\neff$ in the dilute-gas limit (from \cref{eq:neffB2}) obtained by \textit{potter} for the CO$_2$ molecular models and from the Span and Wagner \cite{Span-JPCRD-1996} (S\&W) EOS. \label{fig:B2neff_CO2_models} }
\end{figure}

The molecular models used in this work for CO$_2$ are linear and rigid and do not allow for vibrational contributions to the energy. The dilute-gas limit of $\neff$ can therefore be obtained, as described above, from four-fold integration. Classical values of {\neff} are shown in \cref{fig:B2neff_CO2_models} for the considered molecular models as a function of temperature. The \textit{ab initio} potential of Hellmann \cite{Hellmann-CPL-2014} can yield very accurate predictions of the dilute-gas thermophysical properties (e.g., second virial coefficient). As such, and especially given the physically sound basis of this model, it is believed that the values of $\neff$ from the Hellmann \cite{Hellmann-CPL-2014} model in the dilute-gas limit are therefore a suitable baseline for a comparison with other models. The values of $\neff$ calculated from the model of Hellmann are smaller than those of the other models at all temperatures. The model of Errington (which has a more physically sound exponential repulsion as compared with the $r^{-12}$ repulsion of the other models), is much closer than the other models, which are mostly consistent, but with larger values. The value of $n$ for CO$_2$ proposed in the literature for density scaling \cite{Fragiadakis-PRE-2011} is 13.5 based upon density scaling of shear viscosity data in the liquid phase, which is near the peak value of 13.24 obtained for $\neff$ from the Hellmann \cite{Hellmann-CPL-2014} model.

Next, the values of {\neff} from the molecular models are plotted as a function of temperature and density. Given the qualitative similarities, only the results for the Merker et al. and Hellmann potentials are shown here; the remainder are in the supporting information. Many qualitative features of these results are similar to those of the Lennard-Jones fluid. At high temperatures ($T/T_{\rm crit}\gtrsim 20$), $\neff$ does not change much as the density is swept through a large range, and the temperature at the maximum of $\neff$ along an isochore does not depend strongly on the density; it is close to the maximum obtained from the dilute-gas calculations. The infinite temperature limit for CO$_2$ (unphysically neglecting dissociation) should be 3/2 (see the appendix of Ref. \citenum{Polychroniadou-IJT-2021-Kr}), which holds for all potentials that are finitely valued at all separations. For the Hellmann model, the contributions to the potentials are divergent at a center-of-mass separation of zero, and a small hard core is required for each site-site interaction, which makes the infinite temperature limit go to infinity (see for instance the result for the square-well fluid in Ref. \citenum{Bell-JCP-2020-neff}).

\begin{figure}
	\includegraphics[width=3.5in,page=1]{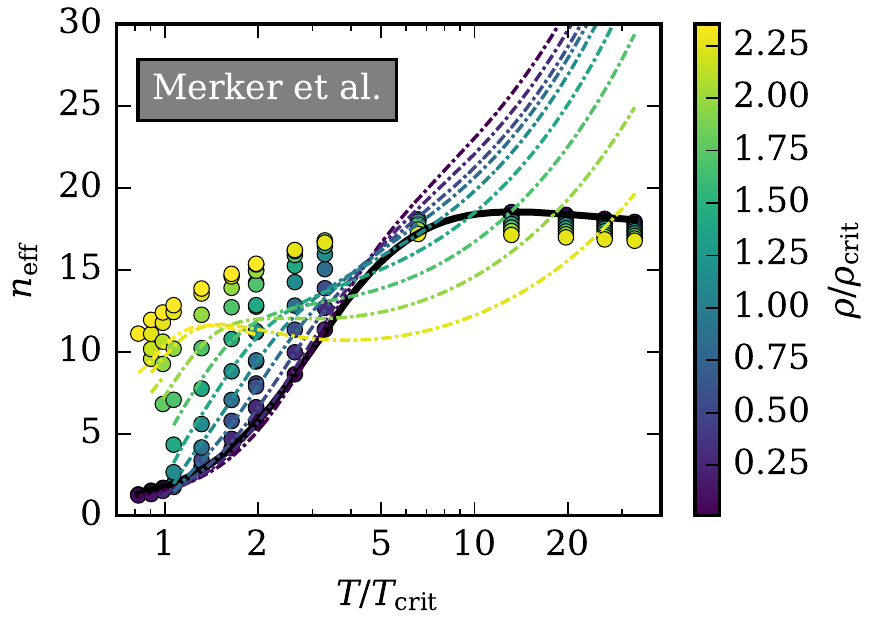}
	\includegraphics[width=3.5in,page=2]{allmodels}
	\caption{Results for $\neff$ for CO$_2$ from the molecular models of \citeauthor{Merker-JCP-2010}\cite{Merker-JCP-2010} and Hellmann \cite{Hellmann-CPL-2014}. The dashed-dotted curve for each isochore is the same quantity given by the Span and Wagner EOS \cite{Span-JPCRD-1996}. \label{fig:neffMerkerHellmann}}
\end{figure}

\section{Conclusions}

Density scaling and entropy scaling can be conceptually aligned by considering density scaling as a mapping onto the residual entropy. The optimal value of $n$ to maximize the correlation between $\Gamma^{-1}$ and $s^+$ appears to have some link to the maximum of $\neff$ in the dilute gas limit. For the Lennard-Jones monomer fluid, the maximum is 15.06 (see \cref{fig:LJ_neff_Trho}), and the optimal scaling value is 15.9.  For the Lennnard-Jones dimer fluid, the maximum values is 18.7 (see \cref{fig:2CLJ_neff}), and the optimal scaling value is 17.5. For the Hellman model for CO$_2$, the maximum value is 13.24 and the optimal scaling value is 12.1.  All scaling values are approximately within one unit of the maximum value. This preliminary observation should be further studied in order to understand whether this relationship should be expected to hold in general. If so, it could offer a route to an entirely predictive approach for entropy scaling that does not require an equation of state or Monte Carlo methods.

The persistent challenge of both density scaling and entropy scaling is that \textit{a priori} predictions of the functional form of the relationship between the scaled transport properties and the independent variable remain out of reach. The hope is that these observations about the relationship between density scaling and entropy scaling might allow for a new approach bearing more fruit. For instance, it was observed for the Lennard-Jones monomer fluid, and indeed for many other fluids, that there is an approximately exponential relationship between macroscopically scaled viscosity times $s^+$ to the power of 2/3 and the residual entropy \cite{Bell-JPCB-2019-LJ}.

\section{Supplementary Material}
	The supplementary material includes a PDF with:
    \begin{itemize}
    	\item Figures like \cref{fig:srMerkerHellmann}, \cref{fig:srdevsMerkerHellmann}, and \cref{fig:neffMerkerHellmann} for the other CO$_2$ models
    	\item Results on change of entropy upon dimerization
    	\item Python snippet for data processing
    	\item Critical region analysis
    \end{itemize}
	The complete set of molecular simulation results for CO$_2$, Lennard-Jones monomer, and Lennard-Jones dimer models are provided in a zip archive.

\begin{acknowledgments}
	Thanks are given to Robert Hellmann for discussion of the relative impact of two- and three-body interactions and Tae-Jun Yoon for discussions of two-body residual entropy. This work was partially supported by the Deutsche Forschungsgemeinschaft (DFG) under grant no. VR 6/16. The simulations were carried out on the Oculus cluster at the University Paderborn and the supercomputer Hawk at the High Performance Computing Center Stuttgart (HLRS) within Project no. MMHBF2. This work was partially supported by the VILLUM Fundation’s \textit{MATTER} Grant (No. 16515).
\end{acknowledgments}

\section{Author Declarations}
\subsection{Conflict of Interest}
The authors have no conflicts to disclose

\subsection{Data Availability}
Data available in article, supplementary material, and linked code repositories. For more detailed information, or additional clarifications, please contact the corresponding author.
    
%


\end{document}